\documentclass[twocolumn,preprintnumbers,amsmath,amssymb,prb]{revtex4}

\usepackage{graphicx}% include figure files
\usepackage{dcolumn}% align table columns on decimal point
\usepackage{bm}% bold math

\begin{document}

\newcommand{\newc}{\newcommand}
\newc{\mbf}{\mathbf}
\newc{\boma}{\boldmath}
\newc{\beq}{\begin{equation}}
\newc{\eeq}{\end{equation}}
\newc{\beqar}{\begin{eqnarray}}
\newc{\eeqar}{\end{eqnarray}}
\newc{\beqa}{\begin{eqnarray*}}
\newc{\eeqa}{\end{eqnarray*}}

\newc{\bd}{\begin{displaymath}}
\newc{\ed}{\end{displaymath}}

%\preprint{aps/123-qed}

\title{Virtual Displacement in Lagrangian Dynamics}

\author{Subhankar Ray}
\email{subho@juphys.ernet.in}
\affiliation{Dept of Physics, Jadavpur University, Calcutta 
700 032, India}
\affiliation{C. N. Yang Institute for Theoretical Physics, 
Stony Brook, NY 11794}
\author{J. Shamanna}
\email{jshamanna@rediffmail.com}
\affiliation{Physics Department, Visva Bharati University, 
Santiniketan 731235, India}

\date{September 1, 2003}% it is always \today today

\begin{abstract}
The confusion and ambiguity encountered by students, in 
understanding virtual displacement and virtual work, is
addressed in this article.
A definition of virtual displacement is presented that allows
one to express them explicitly for both time independent and
time dependent constraints. 
It is observed that for time independent constraints the virtual
displacements are the displacements allowed by the constraints.
However this is not so for a general time dependent case. For simple 
physical systems, it is shown that, the work done on virtual 
displacements by the constraint forces is zero in both the situations. 
For allowed displacements however, this is not always true.
It is also demonstrated that when constraint forces do zero work on
virtual displacement, as defined here, we have a solvable mechanical
problem. We identify this special class of constraints, physically
realized and solvable, as {\it the ideal constraints}.
The concept of virtual displacement and the principle
of zero virtual work by constraint forces are central to both
Lagrange's method of undetermined multipliers, and Lagrange's
equations in generalized coordinates.
\end{abstract}

%\pacs{valid pacs appear here}
% pacs, the physics and astronomy classification scheme.
%\keywords{suggested keywords}%use showkeys class option if keyword
\maketitle

\section{Introduction}
Almost all graduate level courses in classical mechanics
include a discussion of virtual displacement 
\cite{goldstein,sommer,hylleraas,greenwood,schaum,symon,sygr,
taylor,ahaas,terhaar,hand} and
Lagrangian dynamics 
\cite{goldstein,sommer,hylleraas,greenwood,schaum,symon,sygr,
taylor,ahaas,terhaar,hand,landau,arnold}. From the concept 
of zero work by virtual 
displacement the Lagrange's equations of motion are derived. 
However, the definition of virtual displacement is
rarely made precise and often seems vague and ambiguous to students.
In this article we attempt a more systematic and precise 
definition, which 
not only gives one a qualitative idea of virtual displacement,
but also allows one to quantitatively express the same for any given 
constrained system. 
We show that in a number of natural systems, e.g., particle 
moving on a frictionless slope, pendulum with moving point of 
suspension, the work done by the 
forces of constraint due to virtual displacement is zero. 
We also demostrate that this condition is necessary for the
solvability of a mechanical problem.
Hence we propose such systems as an important
class of natural systems.

\subsection{Ambiguity in virtual displacement}
In the following we try to classify the difficulties faced
by a student in understanding the definition of virtual displacement.

\begin{enumerate}
%
% obeys the constraint equations
%
\item It is claimed that (i){\it a virtual displacement $\delta\mbf{r}$
is consistent with the forces and constraints imposed on the system at
a given instant $t$} \cite{goldstein}; 
(ii) {\it a virtual displacement is an arbitrary, instantaneous, 
infinitesimal change of position of the system compatible with the
conditions of constraint} \cite{sommer};
(iii) {\it virtual displacements are, by definition, arbitrary displacements
of the components of the system, satisfying the constraint} \cite{hylleraas};
(iv) {\it virtual displacement does not violate the constraints} \cite{taylor}; 
(v) {\it we define a virtual displacement as one which does not violate
the kinematic relations} \cite{terhaar};
(vi) {\it the virtual displacements obey the constraint on the motion}
\cite{hand}.
These statements imply that the virtual displacements satisfy the constraint 
conditions, i.e., the constraint equations.
However this is true only for time independent (sclerenomous) constraints. 
We shall show that for time dependent (rheonomous) constraints, 
such as a pendulum with moving support, this definition would violate 
the zero virtual work condition.
%
% time passage is zero dt = 0
%
\item It is also stated that (i){\it virtual displacement is to be
distinguished from an actual displacement of the system occurring in a 
time interval $dt$} \cite{goldstein};  
(ii) {\it it is an arbitrary, instantaneous, 
change of position of the system} \cite{sommer};
(iii) {\it virtual displacement $\delta\mbf{r}$ takes place without any 
passage of time} \cite{taylor}.
(iv) {\it virtual displacement has no connection with the time - in
contrast to a displacement which occurs during actual motion, and which
represents a portion of the actual path} \cite{ahaas};
(v) one of the requirements on {\it acceptable virtual displacement
is that the time is held fixed} \cite{hand}.
We even notice equation like `$\delta x_i = d x_i$ for $dt=0$' \cite{taylor}.
The above statements are puzzling to a student.
If position is a continuous function of time, a change in position during 
zero time has to be zero. In other words, this definition implies that 
the virtual displacement cannot possibly be an infinitesimal (or 
differential) of any continuous function of time. 
In words of Arthur Haas: {\it since its} (virtual displacement)
{\it components are thus not fucntions of the time, we are not able to
regard them as differentials, as we do for the components of the element
of the actual path} \cite{ahaas}.
We shall show that virtual displacement can be looked upon as a 
differential, it is indeed a differential increment in virtual 
velocity over a time $dt$, Eq.(\ref{vdisp_time}).
%
% does not obey the constraint equations
%
\item It is also stated that 
(i) {\it virtual displacements do not necessarily conform to the 
constraints} \cite{greenwood};
(ii) {\it the virtual displacements $\delta q$ have nothing to do with
actual motion. They are introduced, so to speak, as test quantities,
whose function it is to make the system reveal something about its
internal connections and about the forces acting on it} \cite{sommer};
(iii) {\it the word ``virtual'' is used to signify that the displacements
are arbitrary, in the sense that they need not correspond to any actual
motion executed by the system} \cite{hylleraas};
(iv) {\it it is not necessary that it} (virtual displacement) {\it 
represents any actual motion of the system} \cite{symon}; 
(v) {\it it is not intended to say that such a displacement} (virtual)
{\it occurs during the motion of the particle considered, or even that
it could occur} \cite{ahaas};
(vi){\it virtual displacement is any arbitrary infinitesimal 
displacement not necessarily along the
constrained path} \cite{schaum}. 
From the above we understand that the virtual displacemnts do not always satisfy 
the constraint equations, and they need not be the ones actually realized.
We shall see that these statements are consistent with physical 
situations, but they cannot serve as a satisfactory definition of virtual 
displacement.
Statements like: ``not necessarily conform to the constraints'' or 
``not necessarily along the constrained path'' only tell us what 
virtual displacement is not, they do not tell us what it really is.
Reader should note that there is a conflict between the claims under 
items 1 and 3. It is not clear from the above, 
whether the virtual displacements
satisfy the constraints, i.e., the constraint equations or not.
\item Virtual displacement is variously described as: {\it arbitrary},
{\it virtual}, and {\it imaginary} \cite{goldstein,sommer,
hylleraas,symon,schaum}. These adjectives make the definition somewhat 
mysterious to a student.
\end{enumerate}

Together with the above ambiguities, students are often confused as
to whether it suffices to understand virtual displacement
as an abstract concept, or they need to have a quantitative definition.
Some students appreciate that the virtual displacement as a vector
should not be ambiguous. The principle of zero virtual work is used 
to derive Lagrange's equations. For a particle under constraint this 
means that the virtual displacement is always orthogonal to the 
force of constraint.

At this stage a student may get further puzzled. Should he take the
forces of constraint as supplied, and then the principle of
zero virtual work as a definition of virtual displacement ? 
In that case the principle reduces merely to a definition of a new 
concept, namely virtual displacement. 
Or should the virtual displacement be defined
from the constraint conditions independently ?
The principle of zero virtual work may then be used to calculate
the forces of constraint that ensure constraint condition 
throughout the motion.

\section{Virtual displacement and Forces of Constraint}
\subsection{Constraints and Virtual displacement}
Let us consider a system of constraints
that are expressible as equations involving positions and time. 
They represent some geometric restrictions (holonomic)
either independent of time (sclerenomous) or explicitly dependent 
on it (rheonomous). Hence for a system of $N$ particles moving in
three dimensions, a system of ($s$) holonomic, rheonomous 
constraints are represented by functions of $\mbf{r}_k$ and ($t$),
\beq{\label{const}}
f_i(\mbf{r}_1,\mbf{r}_2,\dots,\mbf{r}_N,t) = 0,
\hskip 1cm i=1,2,\dots,s
\eeq
Each constraint of this form imposes a restriction on the possible
or {\it allowed velocities}, which must satisfy,
\beq{\label{velconst}}
\sum_{k=1}^{N} \left( \frac{\partial f_i}{\partial \mbf{r}_k} 
\right) \cdot \mbf{v}_k + \frac{\partial f_i}{\partial t} = 0,
\hskip 1cm i=1,2,\dots,s
\eeq
It is worth noting at this stage that there are many, in fact 
infinitely many, allowed velocities, since we have imposed only
($s$) number of constraints on ($3N$) scalar components of the 
allowed velocity vectors. An infinitesimal displacement 
over time ($dt$) due to allowed velocities will be called 
the {\it allowed infinitesimal displacement} or simply allowed
displacement.
\beq{\label{alldisp}}
d\mbf{r}_k = \mbf{v}_k dt
\hskip 1cm k=1,2,\dots,N
\eeq
Allowed displacements $d\mbf{r}_k$ together with differential of
time ($dt$) satisfy constraint equations similar to 
Eq.(\ref{velconst}).
\beq{\label{allcons}}
\sum_{k=1}^{N} \left( \frac{\partial f_i}{\partial \mbf{r}_k} 
\right) \cdot d\mbf{r}_k + \frac{\partial f_i}{\partial t} dt = 0,
\hskip 1cm i=1,2,\dots,s
\eeq
As there are many allowed velocities we have many allowed 
infinitesimal displacements. We propose to define 
{\it virtual displacement} as the difference 
between any two such (unequal) allowed displacements,
\beq{\label{virdisp}}
\delta \mbf{r}_k = d\mbf{r}_k - d\mbf{r}'_k,
\hskip 1cm k=1,2,\dots,N
\eeq
This definition is motivated by the possibility of identifying a 
special class of `{\it ideal constraints}' (sec. IIc), and
verifying `{\it the principle of zero virtual work}' in common
physical examples (sec. III).
It may be noted that, by this definition, virtual displacement
$\delta\mbf{r}_k$ is not a change in position in zero time.
It is rather the difference of any two allowed displacements
during a time $dt$. 
\beq{\label{vdisp_time}}
\delta \mbf{r}_k = (\mbf{v}_k - \mbf{v}'_k) dt,
\hskip 1cm k=1,2,\dots,N
\eeq
The difference of two allowed velocities $\widetilde{\mbf{v}}_k
= \mbf{v}_k - \mbf{v}'_k$ may be defined as the virtual velocity.

The virtual displacements thus defined satisfy the homogeneous 
part of the constraint equation Eq.(\ref{allcons}) (i.e., with 
$\partial f_i/\partial t =0$).
\beq{\label{vircons}}
\sum_{k=1}^{N} \frac{\partial f_i}{\partial \mbf{r}_k} 
\cdot \delta\mbf{r}_k = 0,
\hskip 1cm i=1,2,\dots,s
\eeq
The absence of the $(\partial f_i/\partial t)$ in the above 
equation, Eq.(\ref{vircons}), gives the precise meaning to the statement 
that {\it virtual displacements are the allowed displacements in the case
of frozen constraints}. The constraints are frozen in time
in the sense that we make the $(\partial f_i/\partial t)$ term zero,
though the $\partial f_i /\partial \mbf{r}_k$ term
still involves time. In the case of stationary constraints, i.e.,
$f(\mbf{r}_1,\dots,\mbf{r}_N)=0$,
the virtual displacements are identical with 
allowed displacements as $(\partial f_i/\partial t)$ is zero.
\subsection{Existence of forces of constraints}
In the case of an unconstrained system of $N$ particles
described by position vectors ($\mbf{r}_k$) and  velocity
vectors ($\mbf{v}_k$), the motion is governed by Newton's Law,
\beq{\label{newton}}
m_k \mbf{a}_k = \mbf{F}_k (\mbf{r}_l,\mbf{v}_l,t),
\hskip 1cm k, l =1,2,\dots,N
\eeq
where $m_k$ is the mass of the $k$th particle, $a_k$ is
its acceleration and $F_k$ is the total external force acting 
on it. However, for a constrained system, the equations of 
constraint, namely Eq.(\ref{const}),
impose the following restrictions on the allowed accelerations,
\beqar{\label{acccons}}
\sum_{k=1}^{N} \frac{\partial f_i}{\partial \mbf{r}_k} 
\cdot \mbf{a}_k + \sum_{k=1}^{N} \frac{d}{dt}\left(\frac{\partial 
f_i}{\partial \mbf{r}_k}\right) \mbf{v}_k +\frac{d}{dt}\left(
\frac{\partial f_i}{\partial t}\right) = 0, && \nonumber \\
\hskip 1cm i=1,2,\dots,s &&
\eeqar
Given $\mbf{r}_k$, $\mbf{v}_k$ one is no longer free to choose
all the accelerations $\mbf{a}_k$ independently.
Therefore in general the accelerations $\mbf{a}_k$ allowed
by Eq.(\ref{acccons}) are incompatible with Newton's Law,
\bd
m_k \mbf{a}_k = \mbf{F}_k, \hskip 1cm k=1,2,\dots,N
\ed
This implies that during the motion the constraint 
condition cannot be maintained by the external forces alone.
Physically some additional forces, e.g., normal reaction from 
the surface of constraint, tension in the pendulum string, 
come into play to ensure that the constraints are satisfied,
Hence one is compelled to introduce forces of constraints 
$\mbf{R}_k$ and modify the equations of motion as,
\beq{\label{newtonconst}}
m_k \mbf{a}_k = \mbf{F}_k + \mbf{R}_k, 
\hskip 1cm k=1,2,\dots,N
\eeq
Now the problem is to determine the motion of $N$ particles, 
namely their positions ($\mbf{r}_k(t)$), velocities 
($\mbf{v}_k(t)$)
and the forces of constraints ($\mbf{R}_k$), for a given set of
external forces $\mbf{F}_k$, constraint equations
($f_i(\mbf{r}_1,\mbf{r}_2,\dots,\mbf{r}_N,t) = 0$, $i=1,2,\dots,s$)
and initial conditions 
($\mbf{r}_k(0), \mbf{v}_k(0)$).
It is important that the initial conditions are also
compatible with the constraints.

There are a total of ($6N$) scalar unknowns, namely the
components of $\mbf{r}_k(t)$ and $\mbf{R}_k$, connected by
($3N$) equations of motion, Eq.(\ref{newtonconst}), and ($s$) 
equations of constraints, Eq.(\ref{const}). 
For ($6N > 3N + s$) we have an under-determined 
system. Hence to solve this problem we need ($3N-s$) additional
scalar relations.

\subsection{Solvability and ideal constraints}
In simple problems with stationary constraints, e.g., motion on
a smooth stationary surface, we observe that the allowed
displacements are tangential to the surface. The virtual 
displacement being a difference of two such allowed displacements,
is also a vector tangential to it. The force of constraint,
so called `normal reaction', is perpendicular to the surface.
Hence the work done by the constraint forces on allowed as
well as virtual displacement is zero,
\bd
\sum_{k=1}^N \mbf{R}_k \cdot d\mbf{r}_k = 0, \hskip 1cm 
\sum_{k=1}^N \mbf{R}_k \cdot \delta\mbf{r}_k = 0 
\ed
When the constraint surface is in motion, the allowed
velocities, and hence the allowed displacements are no 
longer tangent to the surface (see sec. III). The virtual 
displacement remains tangent to the constraint 
surface. If the forces of constraint can still be assumed
normal to the instantaneous position of the surface,
we have zero virtual work. However note that the work by
constraint forces on allowed displacements is not zero.
\beq{\label{vwork}}
\sum_{k=1}^N \mbf{R}_k \cdot d\mbf{r}_k \neq 0, \hskip 1cm 
\sum_{k=1}^N \mbf{R}_k \cdot \delta\mbf{r}_k = 0 
\eeq
In a number of physically interesting simple problems, such
as, motion of a pendulum with fixed and moving support, 
motion of a particle along a stationary and moving slope, 
we observe that the above interesting relation 
between the force of constraint and virtual 
displacement holds (see sec. III). 
Out of the above $3N$  virtual displacements, only $n=3N-s$ 
are independent. If the ($s$) dependent quantities are expressed 
in terms of remaining $n=3N-s$ independent objects we get 
\beq{\label{vwork2}}
\sum_{j=1}^n \widetilde{R}_j \cdot \delta \widetilde{x}_j = 0
\eeq
where $\widetilde{x}_j$ are the independent components of 
$\mbf{r}_k$. $\widetilde{R}_j$ are the coefficients of 
$\delta \widetilde{x}_j$, and are composed of different $\mbf{R}_k$.
Since the above components of virtual 
displacements $\delta \widetilde{x}_j$ are independent, 
one can equate each of their coefficients to zero 
($\widetilde{R}_j =0$).
This brings in ($3N-s$) new scalar conditions or equations
and the system is solvable (not under-determined)
again.

Thus we find a special class of constraints which is observed 
in nature (see sec. III) and which gives us a solvable system.
We call this special class of constraints, satisfying zero 
virtual work principle by constraint forces, i.e., 
$\sum_k \mbf{R}_k \cdot \delta\mbf{r}_k = 0$,
the {\it ideal constraint}.

Our interpretation of the principle of zero virtual work, as a 
definition of an ideal class of constraints, find support in
Sommerfeld. In his words, ``{\it a general} {\sf postulate} {\it of mechanics:
in any mechanical systems the virtual work of the reactions
equals zero. Far be it from us to want to give a general proof of
this postulate, rather we regard it practically as definition of a} 
{\sf mechanical system} ''.

\section{Examples of virtual displacements}
\subsection{Simple Pendulum with stationary support}
The motion of the pendulum is confined to a plane and the bob
moves at a fixed distance from the point of suspension.
The equation of constraint by Eq.(\ref{const}) therefore is,
\bd
f(x,y,t) \doteq x^2+y^2- r_0^2 = 0
\ed
Whence
\bd
\frac{\partial f}{\partial x} =2x ,\hskip .3cm
\frac{\partial f}{\partial y} =2y ,\hskip .3cm
\frac{\partial f}{\partial t} = 0
\ed
\begin{figure}[h]
\resizebox{1.8in}{!}
{\includegraphics{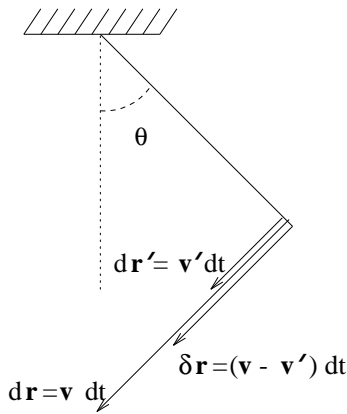}}
\caption{Allowed and virtual displacements for a pendulum with stationary support}
\end{figure}

Hence the constraint equation for allowed velocities (compare
Eq.(\ref{velconst})) is,
\bd
x \cdot v_x + y \cdot v_y =0
\ed
Hence the allowed velocity ($v_x$, $v_y$)
is orthogonal to the instantaneous 
position ($x$,$y$) of the bob relative to stationary support.
The same may also be verified taking a plane polar coordinate.

The allowed displacements are always collinear to allowed 
velocities. Virtual displacement being difference of two 
allowed displacements, is also a vector collinear
to the allowed velocities, hence tangential to the line of
suspension.
\bd
d \mbf{r} = \mbf{v} dt, \;\;\;\; d \mbf{r}' = \mbf{v}' dt
\ed
\bd
\delta \mbf{r} = (\mbf{v} - \mbf{v}') dt
\ed
We may assume that the string of the pendulum provides
a tension ($\mbf{T}$) but no shear (ideal string).
We get zero work by tension due to allowed and virtual
displacements,
\bd
\mbf{T} \cdot d\mbf{r} = 0, \hskip 1cm \mbf{T} \cdot \delta\mbf{r} = 0
\ed
\subsection{Simple Pendulum with moving support}
Let us first consider the case when the support is moving
vertically with a velocity $u$.
The motion of the pendulum is still confined to a plane.
The bob moves keeping a fixed distance from
point of suspension.
The equation of constraint is,
\bd
f(x,y,t) \doteq x^2+(y-u t)^2- r_0^2 = 0
\ed
where $u$ is the velocity of the point of suspension
along a vertical direction.

\begin{figure}[h]
\resizebox{2.0in}{!}
{\includegraphics{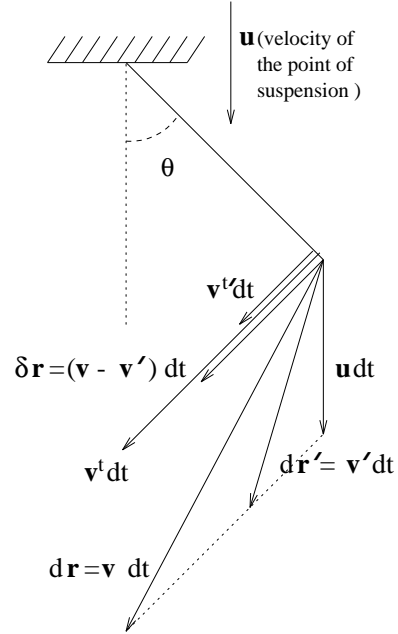}}
\caption{Allowed and virtual displacements for a pendulum with moving support}
\end{figure}

Whence
\bd
\frac{\partial f}{\partial x} =2x ,\hskip .3cm
\frac{\partial f}{\partial y} =2 (y- u t) ,\hskip .3cm
\frac{\partial f}{\partial t} = - 2 u (y - ut)
\ed
Hence the constraint equation gives,
\bd
x \cdot v_x + (y- u t) \cdot v_y - u (y - ut)=0
\ed
or,
\bd
x \cdot v_x + (y- u t) \cdot (v_y - u)  =0
\ed
Hence the allowed velocities ($v_x$, $v_y$)
and hence the allowed displacements,
are not orthogonal to the 
instantaneous position ($x$,$y-ut$) of the bob relative to
the instantaneous position of the support.
It is easy to verify from the above equation that the 
allowed velocity ($v_x$, $v_y$) is equal to 
the sum of a velocity vector ($v_x$, $v_y - u$)
perpendicular to the relative position of the bob with
respect to the point of suspension ($x$,$y-ut$), and the 
velocity of the support ($0$, $u$).
\bd
\mbf{v} = \mbf{v}_{t} + \mbf{u}
\ed

%\begin{figure}[h]
%\resizebox{!}{3.2in}
%{\includegraphics{VirPost/spenVir.ps}}
%\caption{Virtual displacements for a pendulum with moving support}
%\end{figure}

The allowed displacements are vectors collinear to allowed 
velocities. A virtual displacement being the difference of 
two allowed displacements, is a vector collinear to the
difference of allowed velocities. Hence it is tangential 
to the instantaneous line of suspension.
\beqa
d \mbf{r}&=&\mbf{v} dt = \mbf{v}_{t} dt + \mbf{u} dt \\
\delta \mbf{r}&=&(\mbf{v} - \mbf{v}') dt = 
(\mbf{v}_{t} -\mbf{v'}_{t}) dt
\eeqa

At any given instant string provides a tension along its length,
with no shear (ideal string). Hence the constraint force, tension,
still does zero work on virtual displacement.
\bd
\mbf{T} \cdot d\mbf{r} \neq 0, \hskip 1cm \mbf{T} \cdot \delta\mbf{r} = 0
\ed
If one considers the support moving in a horizontal (or in any 
arbitrary direction), one can show that the allowed 
displacement is not normal to the instantaneous line of 
suspension. But the virtual displacement as defined in this 
article always remains perpendicular to the instantaneous 
line of support.
\subsection{Motion along a fixed inclined plane}
The constraint is more conveniently expressed in the polar
coordinate. The constraint equation is,
\bd
f(r,\theta) \doteq \theta - \theta_0 =0
\ed
where $\theta_0$ is a constant. Hence the constraint equation
for allowed velocities, Eq.(\ref{velconst}), gives,
\bd
\sum_{k=1}^{N} \left( \frac{\partial f}{\partial \mbf{r}_k} 
\right) \cdot \mbf{v}_k + \frac{\partial f}{\partial t}
\doteq 
\dot{\theta} + 0 = 0
\ed
Thus the allowed velocities are 
along the constant $\theta$ plane. Allowed velocity,
allowed and virtual displacements are,
\bd
\mbf{v} = \dot{r} \widehat{\mbf{r}}, \hskip .4cm
d\mbf{r} = \dot{r} \widehat{\mbf{r}} dt, \hskip .4cm
\delta\mbf{r} = (\dot{r}-\dot{r}') \widehat{\mbf{r}} dt
\ed
\begin{figure}[h]
\resizebox{!}{1.9in}
{\includegraphics{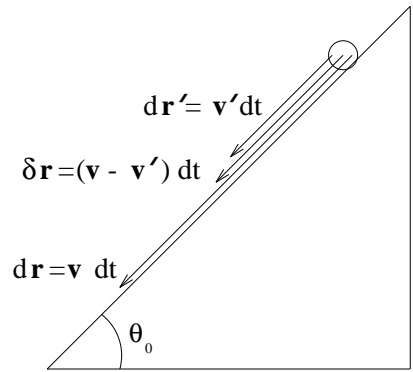}}
\caption{Allowed and virtual displacements for a particle on a stationary slope}
\end{figure}
If the inclined slope is frictionless (ideal), the constraint
force provided by the surface is the normal reaction;
which is perpendicular to the plane.
Hence the work done by this force on allowed as well as virtual
displacement is zero.
\bd
\mbf{N} \cdot d\mbf{r} = 0, \hskip 1cm \mbf{N} \cdot \delta\mbf{r} = 0
\ed
\subsection{Motion along a moving inclined plane}
For an inclined plane moving along the horizontal side,
the constraint is given by,
\beqa
\frac{(x+u t)}{y} - \cot(\theta_0) &=& 0 \\
f(x,y) \doteq (x+u t) - \cot(\theta_0) y &=& 0
\eeqa
whence the constraint for allowed velocities Eq.(\ref{velconst})
become,
\bd
(\dot{x} + u) - \cot(\theta_0) \dot{y} = 0
\ed
Hence the allowed velocity ($\dot{x},\dot{y}$) is the sum of
two vectors, one along the plane ($\dot{x}+u,\dot{y}$), and
the other equal to the velocity of the plane itself ($-u,0$).
\bd
\mbf{v} = \mbf{v}_t + \mbf{u}
\ed

\begin{figure}[h]
\resizebox{!}{2.0in}
{\includegraphics{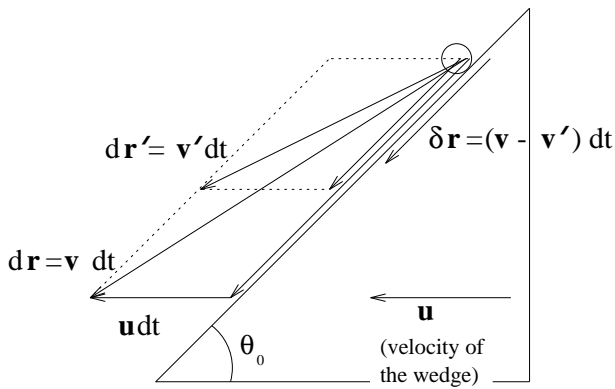}}
\caption{Allowed and virtual displacements for a particle on a moving slope}
\end{figure}
Allowed displacements are vectors along the allowed velocities,
however the virtual displacement is still a vector along the
instantaneous position of the plane.
\beqa
d \mbf{r}&=&(\mbf{v}_t+\mbf{u})dt, \;\;\;\;\;\; d\mbf{r}'=(\mbf{v}'_t +\mbf{u}) dt \\
\delta \mbf{r}&=&(\mbf{v}-\mbf{v}') dt = (\mbf{v}_t-\mbf{v}'_t) dt
\eeqa
For the moving frictionless (ideal) slope, the constraint
force provided by the surface is perpendicular to the plane.
Hence the work done by the constraint force on virtual
displacement is remains zero.
\bd
\mbf{N} \cdot d\mbf{r} \neq 0, \hskip 1cm \mbf{N} \cdot \delta\mbf{r} = 0
\ed
\section{Lagrange's method of undetermined multipliers}
A constrained system of particles follow the equation of
motion given by,
\bd
m_k \mbf{a}_k = \mbf{F}_k + \mbf{R}_k,
\hskip 1cm k=1,2,\dots,N
\ed
where $m_k$ is the mass of the $k$th particle,
$\mbf{a}_k$ is its acceleration. $\mbf{F}_k$
and $\mbf{R}_k$ are the total external force and force
of constraint on the particle.
If the constraints are {\it ideal}, we can write
\beq{\label{vwork00}}
\sum_{k=1}^N \mbf{R}_k \cdot \delta\mbf{r}_k = 0 
\eeq
whence we obtain,
\beq{\label{geneqn}}
\sum_{k=1}^N (m_k \mbf{a}_k - \mbf{F}_k) \cdot 
\delta \mbf{r}_k = 0
\eeq
If the components of $\delta \mbf{r}_k$ were independent, we
could recover Newton's Law for unconstrained system from this
equation. However for a constrained system $\delta \mbf{r}_k$ 
are dependent through the constraint equations,
\beq{\label{const2}}
f_i(\mbf{r}_1,\mbf{r}_2,\dots,\mbf{r}_N,t) = 0,
\hskip 1cm i=1,2,\dots,s
\eeq
or,
\beq{\label{delf}}
\delta f_i = \sum_{k=1}^N 
\frac{\partial f_i}{\partial \mbf{r}_k} \delta \mbf{r}_k = 0,
\hskip 1cm i=1,2,\dots,s
\eeq
We multiply the above equations, Eq.(\ref{delf}), 
successively by $s$ scalar multipliers 
($\lambda_1, \lambda_2, \dots \lambda_s$), 
called the Lagrange's multipliers, and subtract them from 
the zero virtual work equation, Eq.(\ref{vwork00}).
\beq{\label{vwork3}}
\sum_{k=1}^N \left(\mbf{R}_k - \sum_{i=1}^s \lambda_i 
\frac{\partial f_i}{\partial \mbf{r}_k} \right) 
\delta \mbf{r}_k = 0
\eeq
Explicitly in terms of components,
\beq{\label{vwork4}}
\sum_{k=1}^N \left( \left[ R_{k,x} -\sum_{i=1}^s \lambda_i
\frac{\partial f_i}{\partial x_k} \right] \delta x_k
+ [Y]_k \delta y_k + [Z]_k \delta z_k \right) = 0
\eeq
where $[Y]_k$ and $[Z]_k$ denote the coefficients of
$\delta y_k$ and $\delta z_k$ respectively.

The constraint equations Eq.(\ref{const2}) allows us to 
write the ($s$) dependent virtual displacements in terms of 
the remaining $n=3N-s$ independent ones. We choose ($s$) 
multipliers ($\lambda_1,\lambda_2,\dots,\lambda_s$)
such that the coefficients of ($s$) dependent
components of virtual displacement vanish. The remaining
virtual displacements being independent, their coefficients 
must vanish as well. Thus it is possible to choose 
($\lambda_1,\lambda_2,\dots,\lambda_s$)
such that all coefficients ($[X]_k$,$[Y]_k$,$[Z]_k$)
of virtual displacements 
($\delta x_k$,$\delta y_k$,$\delta z_k$) in
Eq.(\ref{vwork4}) vanish. Hence we have the forces of constraint 
in terms of the Lagrange's multipliers.
\beq{\label{fconst}}
\mbf{R}_k = \sum_{i=1}^s \lambda_i 
\frac{\partial f_i}{\partial \mbf{r}_k}, \hskip 1cm k =1,2,\dots, N
\eeq
Thus the problem of mechanics reduces to finding solution of equations
of motion,
\beq
m_k \mbf{a}_k = \mbf{F}_k + \sum_{i=1}^s \lambda_i 
\frac{\partial f_i}{\partial \mbf{r}_k}, \hskip 1cm k=1,2,\dots,N
\eeq
with the constraints,
\beq
f_i(\mbf{r}_1,\mbf{r}_2,\dots,\mbf{r}_N,t)=0, \hskip 1cm i=1,2,\dots,s
\eeq
Thus we have to solve $3N+s$ scalar equations in 
$3N+s$ unknown scalar quantities ($x_k, y_k, z_k,
\lambda_i$). After solving this system we can obtain the
forces of constraint $\mbf{R}_k$ from Eq.(\ref{fconst}).

\section{Generalized coordinates and Lagrange's Equations of motion}
For the sake of completeness we discuss very briefly Lagrange's 
equations in generalized coordinates (for detail see 
\cite{goldstein,sommer,hylleraas,greenwood,schaum,symon,sygr,
taylor,ahaas,terhaar,hand,landau,arnold}).
Consider a system of $N$ particles under $s$ holonomic, rheonomous
constraints of the form given by Eq.(\ref{const}).
We can in principle express $s$ of these coordinates
in terms of the remaining $3N-s$ independent ones.
Or we may express all the $3N$ scalar components of position
in terms of $n=3N-s$ independent parameters $q_1, q_2,\dots,q_n$
and time ($t$).
\beq
\mbf{r}_k = \mbf{r}_k(q_1, q_2, \dots, q_n, t),
\hskip 1cm k=1,2,\dots,N
\eeq
The allowed and virtual displacements are given by,
%differentials, i.e., the differentials of
%position for a fixed `frozen' time t, are the virtual 
%displacements $\delta \mbf{r}_k$:
\beqar
d \mbf{r}_k&=&\sum_{j=1}^n \frac{\partial \mbf{r}_k}
{\partial q_j} \delta q_j+ \frac{\partial \mbf{r}_k}{\partial t}dt, \nonumber \\
\delta \mbf{r}_k&=&\sum_{j=1}^n \frac{\partial \mbf{r}_k}
{\partial q_j} \delta q_j,
\hskip 1cm k=1,2,\dots,N
\eeqar
From the Eq.(\ref{geneqn}) we obtain,
\beq\label{leqn1}
\sum_{k=1}^N m_k \frac{d\dot{\mbf{r}}_k}{dt} \left(\sum_{j=1}^n 
\frac{\partial\mbf{r}_k}{\partial q_j} \delta q_j \right) 
- \sum_{k=1}^N \mbf{F}_k
\left( \sum_{j=1}^n \frac{\partial\mbf{r}_k}{\partial q_j} 
\delta q_j \right) = 0
\eeq
Introducing the expression of kinetic energy,
\bd
T = \frac{1}{2} \sum_{k=1}^N m_k \dot{\mbf{r}}^2_k
\ed
and that of the generalized force,
\beq
Q_j = \sum_{k=1}^N \mbf{F}_k \frac{\partial\mbf{r}_k}{\partial q_j}
\hskip 1cm j=1,2,\dots,n
\eeq
After some simple algebra one finds,
\beq\label{leqn2}
\sum_{j=1}^n \left(\frac{d}{dt}\frac{\partial T}{\partial \dot{q}_j}
- \frac{\partial T}{\partial q_j} - Q_j \right) \delta q_j = 0
\eeq
since the $q_j$ are independent coordinates, coefficient of each 
$\delta q_j$ must be zero separately.
\beq{\label{leqnGF}}
\frac{d}{dt}\frac{\partial T}{\partial \dot{q}_j}
- \frac{\partial T}{\partial q_j} = Q_j,
\hskip 1cm j=1,2,\dots,n
\eeq
In problems where forces $\mbf{F}_k$ are derivable from a 
scalar potential $\widetilde{V}(\mbf{r}_1,\mbf{r}_2,\dots,\mbf{r}_N)$,
\beq
\mbf{F}_k = - \mbox{\boldmath $\nabla$}_k \widetilde{V}(\mbf{r}_1,\mbf{r}_2,
\dots,\mbf{r}_N),
\hskip .6cm k=1,2,\dots,N
\eeq
we can write the generalized force as,
\beq
Q_j = - \mbox{\boldmath $\nabla$}_k \widetilde{V} \cdot
\left( \frac{\partial{\mbf{r}_k}}{\partial q_j} \right)
= - \frac{\partial V}{\partial q_j},
\hskip 1cm j=1,2,\dots,n
\eeq
Where $V$ is the potential $\widetilde{V}$ expressed as a function of 
$(q_1, q_2, \dots, q_n)$.
In addition if the potential $V$ does not depend on
the generalized velocities, we obtain from Eq.(\ref{leqnGF}),
\beq{\label{leqn3}}
\frac{d}{dt} \frac{\partial(T-V)}{\partial \dot{q}_j} - 
\frac{\partial(T-V)}{\partial q_j} = 0, \hskip .6cm j=1,2,\dots,n
\eeq
At this stage one introduces the Lagrangian function $L = T -V$
and in terms of the Lagrangian, the equations of motion Eq.(\ref{leqn3})
take up the form
\beq{\label{leqn4}}
\frac{d}{dt}\frac{\partial L}{\partial \dot{q}_j} -
\frac{\partial L}{\partial q_j} = 0, \hskip .6cm j=1,2,\dots,n
\eeq

\section{Conclusion}
In this article we make an attempt to present a
quantitative definition of the virtual displacement.
We show that for certain simple cases the
virtual displacement does zero work on forces of constraint.
We also demonstrate that this zero work principle allows us
to have a solvable class of problems. Hence we define 
this special class of constraint, {\it the ideal constraint}.
We demonstrate in brief how one can solve a general mechanical 
problem by: i) Lagrange's method of undetermined multiplier and 
ii) Lagrange's equations in generalized coordinates.

In Lagrange's method of undetermined multipliers we have to
solve a larger number ($3N+s$) of equations, than in the case
of Lagrange's equations ($3N-s$) in generalized coordinates.
However we can immediately derive the forces of (ideal) 
constraints in the former case.

It is interesting to note that both the abovementioned
methods require the zero virtual work by constraint forces
as a crucial starting point.
In the case of Lagrange's method of undetermined 
multipliers we start with the ideal constraint condition
Eq.(\ref{vwork00}). From there we write down
Eq.(\ref{geneqn}), Eq.(\ref{vwork3}), Eq.(\ref{vwork4})
and express the constraint forces in terms of Lagrange's 
multipliers, Eq.(\ref{fconst}).
For Lagrange's equations in generalized coordinates
we start with the ideal constraint, Eq.(\ref{vwork00}).
We work our way through Eq.(\ref{geneqn}), Eq.(\ref{leqn1}),
Eq.(\ref{leqn2}) and finally obtain Lagrange's
equations in generalized coordinates, Eq.(\ref{leqnGF})
and Eq.(\ref{leqn4}).

\section*{Acknowledgement}
The authors gratefully acknowledge their teachers in related graduate 
courses at Stony Brook, Prof. Max Dresden, Prof. A. S. Goldhaber and 
Prof. Leon A. Takhtajan.
Authors also acknowledge the encouragement received
from Prof. Shyamal SenGupta of Presidency College, Calcutta. 
The material presented here was used in graduate level 
classical mechanics courses at Jadavpur University during $1998-2001$.
SR would like to thank his students, in particular,
A. Chakraborty (J.U.) for pointing out the 
difficulty in understanding the concept of virtual displacement
in its usual presentation.
Authors have greatly benefited from the books mentioned in this
article, particularly those of Sommerfeld \cite{sommer},
Hylleraas \cite{hylleraas} and Arnold \cite{arnold}.

\vfill


\end{document}